# TEXT MINING TO IDENTIFY AND EXTRACT NOVEL DISEASE TREATMENTS FROM UNSTRUCTURED DATASETS

Rahul Yedida[a], Saad Mohammad Abrar[a], Cleber Melo-Filho[b], Eugene Muratov[b], Rada Chirkova[a], Alexander Tropsha[b]
[a]*Department of Computer Science, North Carolina State University*
[b]*Eshelman School of Pharmacy, University of North Carolina at Chapel Hill*

## ABSTRACT

**Objective:** We aim to learn potential novel cures for diseases from unstructured text sources. More specifically, we seek to extract drug-disease pairs of potential cures to diseases by a simple reasoning over the structure of spoken text.
**Materials and Methods:** We use Google Cloud to transcribe podcast episodes of an NPR radio show. We then build a pipeline for systematically pre-processing the text to ensure quality input to the core classification model, which feeds to a series of post-processing steps for obtaining filtered results. Our classification model itself uses a language model pre-trained on PubMed text. The modular nature of our pipeline allows for ease of future developments in this area by substituting higher quality components at each stage of the pipeline. As a validation measure, we use ROBOKOP, an engine over a medical knowledge graph with only validated pathways, as a ground truth source for checking the existence of the proposed pairs. For the proposed pairs not found in ROBOKOP, we provide further verification using Chemotext
**Results:** We found 30.4% of our proposed pairs in the ROBOKOP database. For example, our model successfully identified that Omeprazole can help treat heartburn.We discuss the significance of this result, showing some examples of the proposed pairs.
**Discussion and Conclusion:** The agreement of our results with the existing knowledge source indicates a step in the right direction. Given the *plug-and-play* nature of our framework, it is easy to add, remove, or modify parts to improve the model as necessary. We discuss the results showing some examples, and note that this is a potentially new line of research that has further scope to be explored. Although our approach was originally oriented on radio podcast transcripts, it is input-agnostic and could be applied to any source of textual data and to any problem of interest.

**Keywords:** information extraction, text mining, knowledge graphs, drug-disease associations

## 1. BACKGROUND AND SIGNIFICANCE

Research on diseases and possible cures is fundamentally restricted by the increasing failure rate, coupled with the lack of monetary value for pharmaceutical companies that might otherwise invest in R\&D for such diseases [1]. Thus, it leads to a lack of research and attention towards associations of diseases that could otherwise hold massive potential. We equate this to the Gordian knot problem, and in this paper, attempt to “cut the Gordian knot [2] by proposing new approaches to research novel cures for diseases.

There were previous attempts to use nontraditional data sources in the medical community. For instance, Mohammadhassanzadeh et al. [3] and Noh et al. [4] used podcasts as knowledge sources. In addition, we note that people, as social beings, tend to discuss various matters with each other;

this is particularly true for associations of diseases. Consequently, this mode of communication naturally lends to a significant amount of data in the form of unstructured natural language for mining useful information. However, this is a fundamentally challenging Natural Language Processing (NLP) task [5]. Recent advances in NLP systems have enabled the training of so-called “language models”, multi-task general-purpose learners that model the structure of natural language sentences and perform various tasks such as part-of-speech tagging [6]. Perhaps the most famous of these language models is the recently proposed BERT model [7] by Google, which achieved state-of-the-art results on several NLP tasks. We instead choose to use a different state-of-the-art language model, FLAIR [8], for the modular nature of its code and its ease of use. Further, the work on BioFLAIR [9] provided clear examples of using the FLAIR model on PubMed, upon which we built our code.

Our research is motivated by the lack of significant research efforts in identifying potential treatments for diseases in non-conventional data sources. Therefore, we attempt to automatically extract knowledge from unstructured text sources to identify potential drug-disease pairs that are plausible treatments. Another constraint here is to involve a human only when necessary, because human experts are typically expensive and busy. In summary, we use a state-of-the-art NLP system as part of a pipeline to close in on plausible treatments for diseases from unstructured text; these can then be validated by human experts, who may decide which are worth pursuing further.

Our motivation for designing a modular system is twofold: first, it is a better practice from a software engineering perspective to create a system whose components are loosely coupled but highly cohesive; this leads to our second, main reason: it allows future researchers to easily improve upon our results by modifying various parts of our pipeline. We fully support open science initiatives and made our source code publicly available as well. We believe this transparent, open-source approach will help promote further work in this area, which is one of our core intents behind this paper.

We would like to summarize the main outcomes of this study below:

- We developed and validated a novel framework for extracting meaningful drug-disease pairs from unstructured, natural language;
- We designed a modular *plug-and-play* system that allows for easily improving upon our system;
- We released our code publicly for reproducibility and allowing future work to build upon our results;
- We emphasized the importance of rigorous validation, our way of performing this, and the hazards of blind trust in knowledge sources.

## 2. DATA AND METHODS

Our approach consisted of three major parts: pre-processing, classification, and post-processing. Figure 1 demonstrates the steps of our pipeline.

### 2.1 The source dataset

The input podcast data consisted of 199 episodes from The People's Pharmacy Radio Program [10]. The raw MP3 files were then passed through Google Cloud to transcribe the episodes to text. The raw result of this is a JSON file for each episode, chunked, with no periods demarcating sentences, where each chunk is associated with a confidence score. On average, each episode had

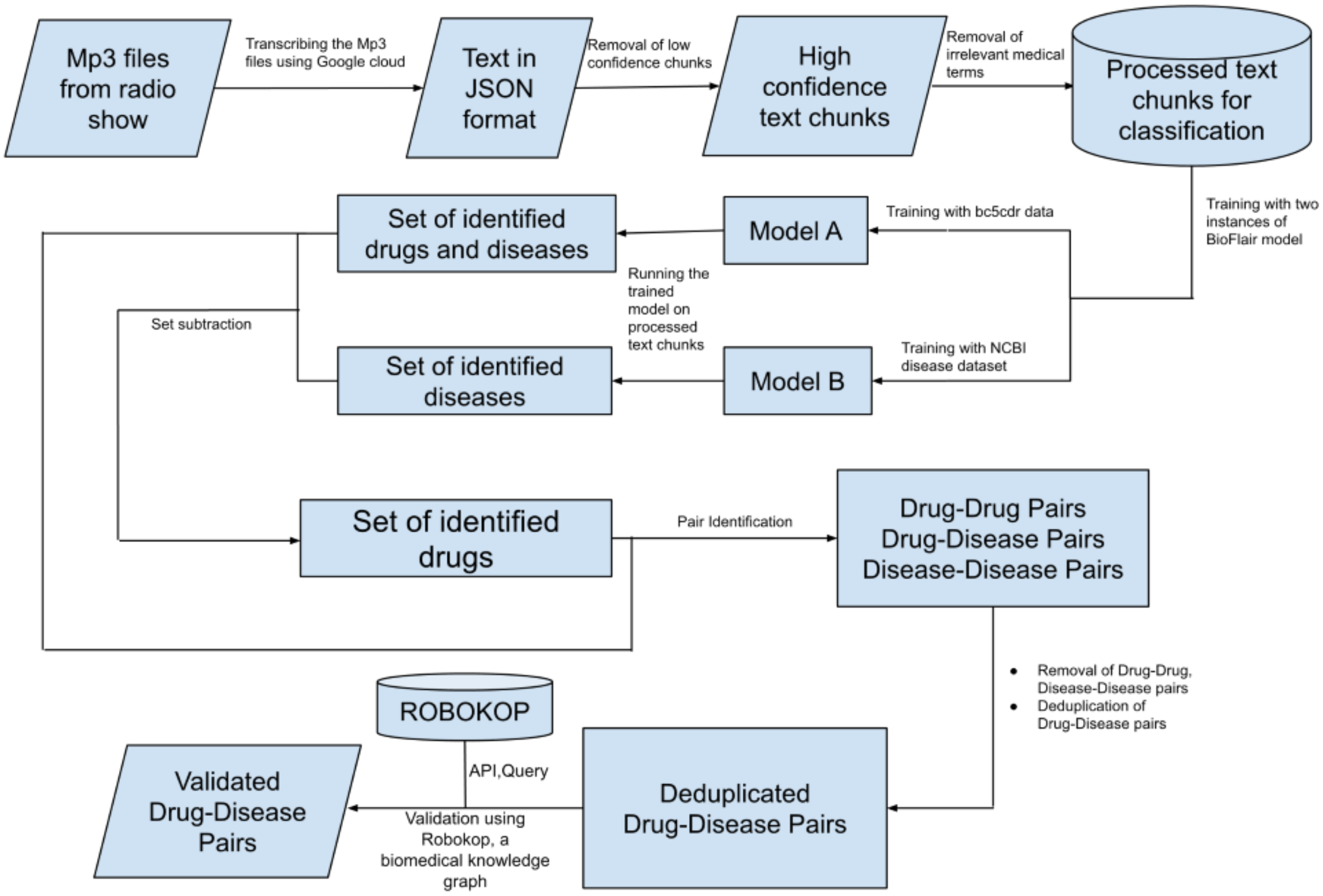

Figure 1: Our pipeline

81 chunks. Figure 2 illustrates the distribution of the number of text chunks in the episodes. In total, we processed 16,291 chunks, some of which were eliminated by our pre-processing steps.

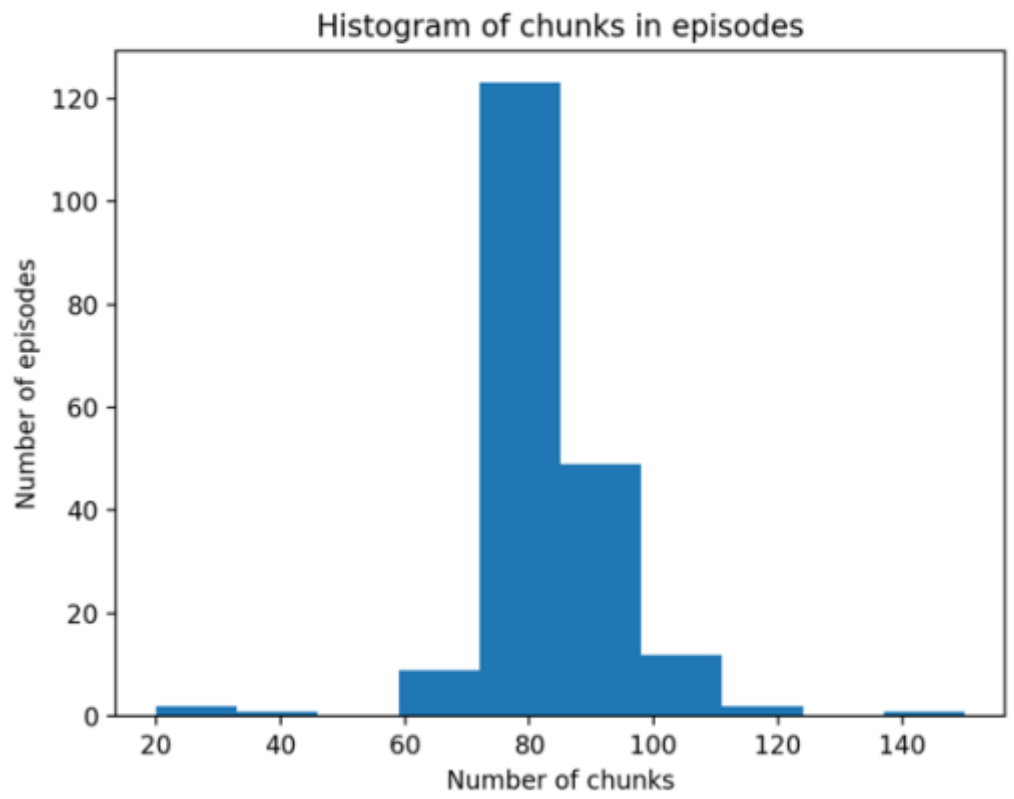

Figure 2: Histogram of number of chunks in podcast episodes.

## 2.2 Pre-processing

Our first pre-processing step (starting from the text data) was removing chunks that have a confidence score less than 96%. We derived this confidence score by manually going through 100 chunks of text and analyzing the degenerate chunks of text (as a particularly egregious example, “Dr. Sacks” was transcribed as “Doctor Sex”).

Table 1: List of medical-adjacent terms

| | | |
|---|---|---|
| pharmacy | health | allergist |
| patient | biochemical | medication |
| symptom | poop[1] | feces |
| urologist | polypharmacy | relief |
| nutrient | dr | herbal |
| care/healthcare | prognosis | diagnosis |
| die/death | insurance | remittance |
| relapse | physician | decease |
| practitioner | anesthesia | ICU |
| Medicare | etiology | |

Because our core idea is to identify neighboring drug and disease terms (together called “medical terms”), we use a language model to identify these medical terms. However, such a language model would also identify “medical-adjacent” terms, such as “doctor” and “hospital”. It is of no value to learn that visiting a doctor helped cure a patient of some disease. Therefore, the next pre-processing step was to compile a list of such medical-adjacent terms. This compiled list is summarized in Table 1. During our pass through the 100 chunks of text, we were able to compile a comprehensive list of these terms. We perform a partial string match and remove all such words from all chunks of text. Because our core idea revolves around pair formation between adjacent medical terms, the semantics of the sentence do not matter, and this does not affect our results.

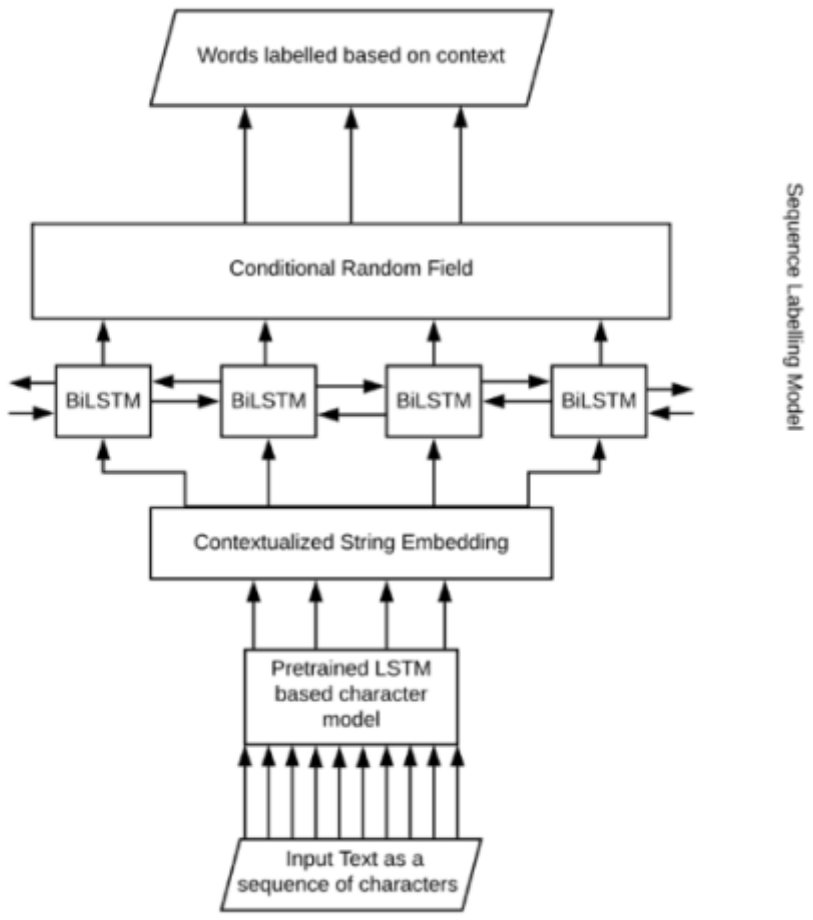

*Figure 3. Architecture of the FLAIR language model*

## 2.3 Classification

[1] This was an actual word in the text. The natural language text is discussed by non-experts, and therefore involves a lot of colloquial terms. Our language model can handle this.

Relevant to our problem statement, we seek out for the model paradigm that adapts the following strategy "one word, multiple embeddings". To be more specific, generating embeddings of a word keeping the latent contextual semantics of the text chunk in mind will generate a more favorable classifier in our case. There has been a breakthrough in contextualized word embedding models and state of the artwork done by Peter et al [6], Chiu et al. [7] have all achieved high F-1 scores in named entity recognition (NER) tasks. Akbik et al. [8] proposed a novel contextualized word embedding model that operates in character level. Taking advantage of a pre-trained character language model, they generated string embeddings carrying contextual meaning. Hence using stacked layers of Bidirectional Long Short Term Memory (BiLSTM), which helps take into consideration an effectively infinite amount of context on both sides of a word, and Conditional Random Field(CRF), they generate a state of the art model for NER tasks that achieves a F-1 score of 93.18% and 89.3% on datasets such as CoNLL-2003 [9] and Ontonotes. The framework, FLAIR [15] is built to accommodate this powerful, state-of the art language model. The overall architecture of the model is illustrated in Figure 3.

Therefore, given the high modularity and accuracy of the FLAIR language model in state-of-the-art datasets, we leverage it to train a model that fits our task. We use the FLAIR language model for its modular nature. At a high level, we use the embeddings trained on PubMed, and pre-train it on our data. Finally, we run Named Entity Recognition on each chunk separately.

In detail, we use two instances of the FLAIR [15] model. We fine-tune one model on the bc5cdr data [17], which contain labels for identifying both diseases and drugs. We fine-tune the other model on the NCBI disease dataset [18], which contains only disease labels. These datasets were obtained from the BioFLAIR [19] code repository. Next, we identify the drugs in text as a set difference between the two models' outputs. We use this twin classification model because it leads to more stable results. Specifically, this approach has two advantages:

- using this model yields less pairs for a human expert to verify;
- for a pair to be proposed, it needs to be acknowledged by two differently trained models; this is similar to asking for two different perspectives from two experts, each of whom studied the same material from different sources.

This robust approach is not very computationally expensive; in our testing, the significant computation (besides the fine-tuning) was in initializing the models, not in classifying. Therefore, our approach leads to a fast, scalable, robust approach to identifying perplexing drug-disease pairs for further examination from a human expert.

### 2.4 Post-processing

Our post-processing steps are designed to work in conjunction with the classification step and prune the pairs proposed. We disregard adjacent medical terms of the same class (i.e., a drug cannot treat another drug; a disease cannot treat another disease), so that the final proposed pairs are at least semantically valid. We also remove adjacent duplicate medical terms. Finally, we, remove duplicate pairs (this includes pairs of the form A, B and B, A).

## 3. RESULTS

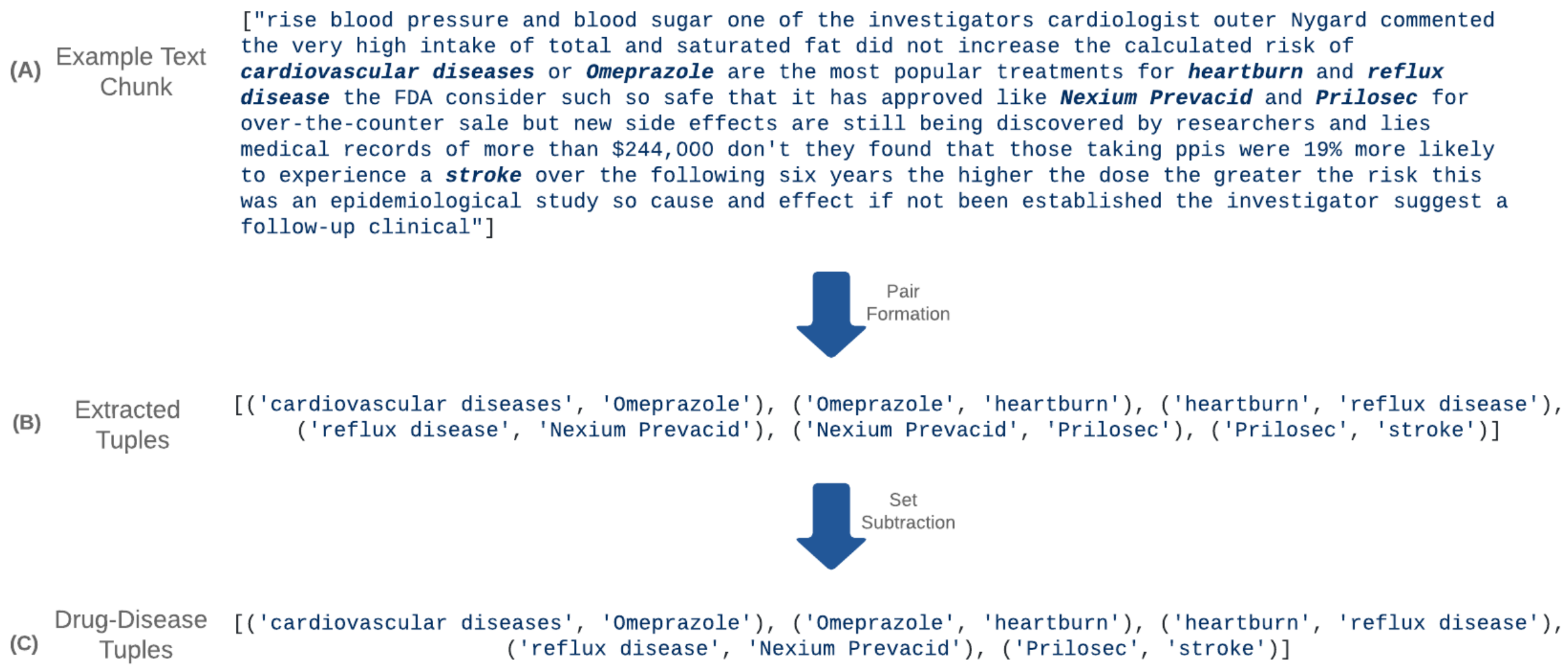

Figure 4: Drug-Disease tuple generation from raw text chunks: An example

Amongst the 16,291 text chunks in a total of 199 episodes, 6514 pairs of disease associations were found by our model. A demonstration of the model working on an example text chunk is provided in Figure 4. As illustrated, the pair formation module creates associations between the adjacent medical terms. Following that, using the set subtraction, only the drug-disease pairs were extracted from the raw data.

Table 2: **Reported Associations of selected disease and verification found in ROBOKOP and Chemotext.** Underlined terms present stronger evidence, i.e., higher number of publications supporting the identified associations.

| Selected Disease | Selected Associations | Verified by ROBOKOP | Verified by ChemoText |
|---|---|---|---|
| Alzheimer's Disease | Resveratrol, Marijuana, Fish Oil,Urea, Beta Amyloid, Aluminum,Coconut oil, Benzodiazepines,Serotonin, Theracurmin, Ketones,herpes simplex virus HSV-1,homocysteine, Pregnenolone,omega-3 polyunsaturated fatty acids,Adenosine Triphosphate,Estradiol, Vitamin D, Ginkgo,Alcohol, Copper,Cholesterol,Zolpidem, Capsaicin, Magnesium,P Gingivalis, Glucose, Curcumin,Aspirin, Epilepsy, Choline | No matches | Resveratrol, Marijuana, Fish Oil, Coconut oil, Benzodiazepines, Ketones, herpes simplex virus HSV-1, Pregnenolone, polyunsaturated fatty acids, Ginkgo, Zolpidem, Capsaicin, Curcumin |
| Asthma | Latanoprost, Vitamin D, Fructose, Marijuana, Magnesium, Vitamin C, Steriods, Corticosteroid, | Pain, Infection, Obesity, Inflammation, Histamine, Diabetes, Epinephrine, Sneezing, | Latanoprost, Fructose, Airway Remodeling |
| | Infection, Airway Remodeling Pain, Obesity, Inflammation, Histamine, Diabetes, Epinephrine, Sneezing, Autoimmunity, Allergic, Progesterone, Timolol, Fluticasone | Autoimmunity, Allergic, Progesterone, Timolol, Fluticasone | |
| Arthritis | Tetracycline, Doxycycline, Inflammation, Steriod, Diclofenac, Ibuprofen, Confusion, Calcium, Curcumin, Naproxen, Weight Loss, Cortisone, Cataract, Magnesium, Meloxicam, Dopamine, Warfarin, Cannabidiol, Tylenol, Vioxx Celebrex, Turmeric, Alpha-gal meat allergy, Boswellia, Dextrose, Parovirus B19 | Tetracycline, Doxycycline, Inflammation, Steriod, Diclofenac, Ibuprofen, Confusion, Calcium, Curcumin, Naproxen, Weight Loss, Cortisone, Cataract, Magnesium, Meloxicam, Dopamine | Warfarin, Cannabidiol, Celebrex, Tylenol |
| Diabetes | Aloe Vera, Alpha Lipoic Acid, Amaryl, Vitamin B12, Cinnamon, Coffee, Confusion. Corticosteroids, Galega Officinalis, NSAID, oligo monosaccharides, oxygen, Viagra, sodium preservative heme iron,White wine, ACE, Alcohol, Anxiety, Asthma, Blindness, Calcium, Codeine, Depression, Fatty Liver, Fibrate, Flu, Fractures, Obesity, Paroxetine, Statin, Succinylcholine, Trauma, Urea, Vitamin D, Weight Loss, Hypoglycemic, Magnesium, Insulin Resistance, Zinc, Glucose, Polycystic Ovary Metformin | ACE. Alcohol, Anxiety, Asthma, Blindness, Calcium, Codeine. Depression, Fatty Liver, Fibrate, Flu, Fractures, Obesity, Paroxetine, Statin, Succinylcholine, Trauma, Urea, Vitamin D, Weight Loss, Hypoglycemic, Magnesium, Insulin Resistance, Zinc, Glucose, Polycystic Ovary Metformin | No Matches |
| Dementia | Arrythmia. Cognitive decline, Anxious, Antidepressants, Omeprazol, Lansoprazol, Ashwagandha, benzodiazepine, Uninary Incontinence stretch, PPI, ulcers, smoking, Aspirin, Antihistamines, Zika Virus, Urea, Glyphosate, Selenium, Prilosec Prevacid | Cholesterol, Fatty Acids, Hyperactivity, Fracture, Curcumin, Sodium. Statin, Nitrate, Agitation, Vitamin B12, Vitamin E, Neurodegeneration, Amyloid, Acetylcholine, Beta-Amyloid, Valacyclovir, Lithium, Lead, Kidney Disease | Arrythmia, Antidepressants, benzodiazepine, smoking[2] |
| Breast Cancer | Vitamin D, Progestin, Estrogen, Multivitamins, Omeprazole, Lansoprazole, Esomeprazole, HRT (Hormone | Bisphenol-a, Lignan | Omeprazole, Lansoprazole, Esomeprazole, Pyrethroids |

[2] one study associating involuntary smoking with dementia

| | | |
|---|---|---|
| | Replacement Therapy), Melanoma, Metformin, Herceptin, Cocoa Flavonoids, DCIS (Ductal Carcinoma in Situ), Dexlansoprazole, Non-Hodgkin’s lymphoma, Pyrethyroids, Dexlansoprazole, phthalates, testosterone, Bisphenol-a, Lignan | |

Owing to the fact that several text chunks reported the same kind of disease associations, several duplicated pairs were reported. Hence, as a part of the post-processing step, deduplication was done. Following the deduplication, the new number of disease associations amounted to 4283. The deduplicated disease associations were checked for validation by querying the ROBOKOP knowledge graph and a total of 976 pairs were verified. Therefore around 22.8\% recall was achieved. The comparative counts of the pairs are reported in Figure 5.

Although the results indicate that the model was able to eke out meaningful pairs of disease associations, there are a number of unverified associations. Thus as an additional validation step, we submitted the associated terms predicted, but not identified by ROBOKOP, to Chemotext [20], a webserver for mining publications in PubMed according to Medical Subject Headings (MeSH) terms. The results for selected diseases occuring in the podcast, the reported associations of the diseases and the verified associations are reported in Table 2.

### 3.1 Examples of associations found in the transcripts and confirmed by Chemotext

The list of associated terms confirmed by Chemotext is available in Table 2. The underlined associations in the table indicated a stronger association in the literature. Some examples of confirmed associations are presented below.

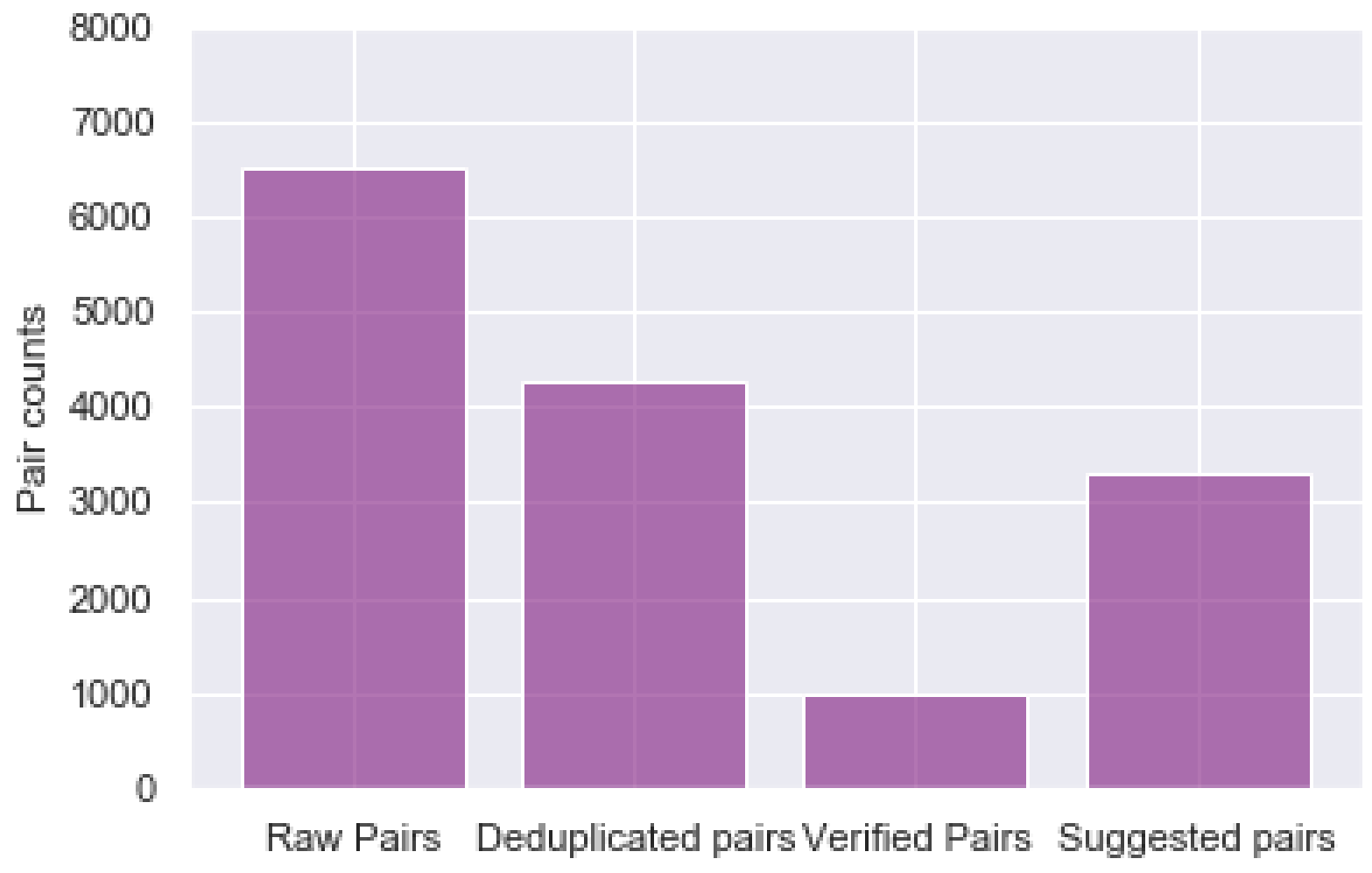

Figure 5: Comparative counts of Raw and Post-processed Pairs

3.1.1 Alzheimer's Disease and benzodiazepines

Benzodiazepines are commonly used for the treatment of certain symptoms of Alzheimer's Disease (AD) [21]. However, pharmacokinetic drug-drug interactions between benzodiazepines and donepezil, an anti-AD drug, have been described [22]. These drugs are metabolized by CYP3A4 and compete for the same catalytic site. Consequently, the rate of transformation of donepezil is reduced if concomitantly used with benzodiazepines [22]. The co-occurrence of the terms "Alzheimer's Disease", "benzodiazepines", and "olanzapine", a thienobenzodiazepine from the class of atypical antipsychotic drugs, was observed in 62 publications. Olanzapine, among other atypical antipsychotics, is used for the treatment of behavioral disturbances in AD patients [23]. However, as some studies suggest, the use of atypical antipsychotics should be closely monitored due to severe side effects, e.g., aggravation of cognitive dysfunction [24] and increase of mortality rate in AD patients [25]. Additional studies have demonstrated that olanzapine dangerous effects for AD are related to downregulation of THRA and PRNP, and upregulation of IL1A expression [26].

3.1.2 Arthritis and Celebrex (celecoxib)

The search in Chemotext returned 69 publications containing the pair "Arthritis/Celebrex" concurrently with the term "Non-Steroidal Anti-Inflammatory Agents". Celecoxib, commercially known as Celebrex, is a nonsteroidal anti-inflammatory drug (NSAID), selective inhibitor of cyclooxygenase-2 (COX-2), commonly used to treat pain in patients with arthritis [27]. Despite being largely used, celecoxib can lead to long-term harmful side effects in patients with arthritis [28]. There is evidence that celecoxib may increase the risk of atrial fibrillation in arthritis patients as evidenced by the increased duration of the P-Wave, an electrophysiological marker for arrhythmia [29]. Because COX-2 is constitutively expressed in the kidneys, other point of concern should be renal toxicity. Therefore, concomitant administration of celecoxib and angiotensin-converting enzyme (ACE) inhibitors should be avoided in patients with compromised renal function due to increased risk of renal insufficiency [27].

3.1.3 Breast cancer and pyrethroids

Synthetic pyrethroids are one of the most broadly used pesticides and in vitro studies have demonstrated their potential estrogenic activity, mainly through their metabolites, by inducing MCF-7 human carcinoma breast cells proliferation [30]. However, no estrogenic activity was observed in orthogonal in vitro and in vivo studies with the metabolites 3-phenoxybenzoic acid and 3-phenoxylbenzil alcohol [31]. Bifenthrin (BF), a pyrethroid insecticide, has demonstrated a potential enantiomer-dependent estrogenic activity by inducing in vitro MCF-7 cells proliferation and vitellogenin induction in aquatic vertebrates [32]. In another study, the pyrethroids fenvalerate and permethrin showed antiprogestagenic in T47D human breast cancer cells [33]. Regardless of how these substances modulate estrogenic and progestogenic activities, their potential for endocrine disruption is undoubtedly a matter of concern.

## 4. DISCUSSION

We discuss some of the aspects and potential concerns with our method and seek to provide additional reasoning for our approach in this section.

### 4.1 Text chunking

It may be noted that eliminating some of the chunks adds discontinuity to the text. However, we argue that this should not significantly impact our results; rather, that it helps generate more robust results. By eliminating chunks below the threshold, we start with better, more robust data. We argue that this makes it less of a burden on the rest of the process to filter out bad results.

Further, by not running our model on two continuous chunks, we do not use a mix of two potentially discontinuous chunks of text as input, which could lead to erroneous results.

### 4.2 Irrelevant term removal

Simply removing the irrelevant terms from the text may cause the resulting text to become ungrammatical and incoherent. However, because our approach does not have a strong reliance on the structure of the sentence, this does not affect our results. Specifically, our approach is based on adjacent medical term pair formation; by removing potentially misleading terms, which we call medical-adjacent terms, we remove potential bad pairs, leading to robust results.

### 4.3 Pair formation

Because the BioFLAIR classifier identifies both diseases and treatments as medical terms, it is certainly possible to have pairs formed consisting of only diseases or only treatments. To alleviate this, we used a twin classifier model in the classification and pair formation stage. We train one BioFLAIR model as described above, and another that identifies only diseases, by training on a disease-only dataset. Then, we form pairs only from the set difference of the terms detected by these models. We find this significantly reduces (by 64.5%) the number of pairs suggested by the overall pipeline, and leads to better that may improve the results significantly.

### 4.4 Validation

The validation step has been a huge concern given the forbidden number of pairs generated to be manually checked. Although, ROBOKOP provided with a platform for validation of possible associations. However, some true positives given by the model are not present in the underlying ROBOKOP Knowledge Graph. These absences, rather than indicating incorrectness, shows new possible associations that were otherwise unknown. Therefore, further associations were validated using Chemotext, to prove the efficiency of the model to report new pairs. In addition to that, there were also some false positives found in the ROBOKOP Knowledge Graph, where trivial relationships that are found by the model (but do not represent meaningful disease-treatment) pairs are marked as correct.

## 5. CONCLUSION

The paper aims at constructing an automated pipeline to identify drug-disease tuples from unstructured texts and we have experimented using the radio podcasts dataset. While the current pipeline achieves a 22.8% recall on the ROBOKOP knowledge graph, there is certainly scope for improvement, and the validation using Chemotext proved out to be fruitful as there were many other associations missed out by ROBOKOP. Therefore, an avenue of future work could involve a more thorough validation using Chemotext and other existing biomedical knowledge sources. Another possible method of computing the true rates may involve random sampling from the results, asking humans for input, and then statistically estimating the true false positive and false negative rates.